# Control of thermoelectric properties of phase-coherent molecular wires

Víctor M. García-Suárez[a,b], Colin J. Lambert[a]*, David Zs. Manrique[a] and Thomas Wandlowski[c]

[a]Department of Physics, Lancaster University, Lancaster LA1 4YB, United Kingdom, c.lambert@lancaster.ac.uk

[b]Departamento de Física, Universidad de Oviedo and CINN (CSIC), ES-33007 Oviedo, Spain

[c]Department of Chemistry and Biochemistry, University of Bern, Freiestrasse 3, 3012 Bern

E-mail: C.lambert@lancaster.ac.uk

*KEYWORDS: OPE, molecular wires, single molecule conductance, thermopower, STM break junction (STM-BJ), mechanically controlled break junction (MCBJ), ab-inito transport calculations, density functional theory, non-equilibrium Green's functions, quantum interference.*

**Abstract:**

We demonstrate how redox control of intra-molecular quantum interference in phase-coherent molecular wires can be used to enhance the thermopower (Seebeck coefficient) $S$ and thermoelectric figure of merit $ZT$ of single molecules attached to nanogap electrodes. Using first principles theory, we study the thermoelectric properties of a family of nine molecules, which consist of dithiol-terminated oligo(phenylene-ethynylenes) (OPEs) containing various central units. Uniquely, one molecule of this family possesses a conjugated acene-based central backbone attached via triple bonds to terminal sulfur atoms bound to gold electrodes and incorporates a fully conjugated hydroquinonecentral unit. We demonstrate that both $S$ and the electronic contribution $Z_{el}T$ to the figure of merit $ZT$ can be dramatically enhanced by oxidizing the hydroquinone to yield a second molecule, which possesses a cross-conjugated anthraquinone central unit. This enhancement originates from the conversion of the pi-conjugation in the former to cross-conjugation in the latter, which promotes the appearance of a sharp anti-resonance at the Fermi energy. Comparison with thermoelectric properties of the remaining seven conjugated molecules demonstrates that such large values of $S$ and $Z_{el}T$ are unprecedented. We also evaluate the phonon contribution to the thermal conductance, which allows us to compute the full figure of merit $ZT = Z_{el}T/(1 + \kappa_p/\kappa_{el})$, where $\kappa_p$ is the phonon contribution to the thermal conductance and $\kappa_{el}$ is the electronic contribution. For unstructured gold electrodes, $\kappa_p/\kappa_{el} \gg 1$ and therefore strategies to reduce $\kappa_p$ are needed to realise the highest possible figure of merit.

TEXT:

# 1. Introduction.

Generation of electricity from heat via the Seebeck effect is silent, environmentally friendly and requires no moving parts. Waste heat from automobile exhausts and industrial manufacturing processes could all be used to generate electricity provided materials with a high thermoelectric figure of merit *ZT* could be identified. Conversely Peltier cooling using such materials would have applications ranging from on-chip cooling of CMOS-based devices to home refrigerators. Despite several decades of development, the best inorganic thermoelectric materials, such as bismuth telluride (Bi2Te3)-based alloys, possess a figure of merit *ZT* close to unity only, which is not sufficient to create a viable technology platform for harvesting waste heat. As an alternative, organic thermoelectric materials are now being investigated and are already showing promising values of both *ZT* and thermopower (Seebeck coefficient)[1].

The thermopower or Seebeck coefficient (*S*) and thermoelectric figure of merit (*ZT*) of a material or of a nanojunction are defined as $S = -\Delta V/\Delta T$ and $ZT = S^2 GT/\kappa$, where $\Delta V$ is the voltage difference created between the two ends of the junction when a temperature difference $\Delta T$ is established between them, *G* is the electrical conductance, *T* is the ambient temperature and *κ* is the thermal conductance. A key strategy for improving the thermoelectric properties of inorganic materials has been to take advantage of nanostructuring[2-4], which leads to quantum confinement of electrons and enhanced thermoelectric performance[4-10]. The single-molecule building blocks of organic materials offer the ultimate limit of electronic confinement, with quantised energy level spacings, which are orders of magnitude greater than room temperature. Therefore it is natural to examine the thermoelectric performance of single-molecule junctions as a stepping stone towards the design of new materials. The ability to measure thermopower in single-molecule junctions is relatively new[11-16] and the thermoelectric properties of only a few molecules have been measured. To underpin further exploration and optimisation of single-molecule thermoelectric devices, the aim of this paper is to survey theoretically the thermoelectric performance of a family of related single molecules and demonstrate that redox control is a viable strategy for enhancing both *S* and *ZT*.

Initially, when discussing the figure of merit, we shall focus on the electronic contribution $Z_{el}T$ to $ZT$, defined by $Z_{el}T = S^2 GT/\kappa_{el}$, where $\kappa_{el}$ is the electronic contribution to the thermal conductance. The full figure of merit is then given by $ZT = Z_{el}T /(1 + \kappa_p/\kappa_{el})$, where is $\kappa_p$ the phonon contribution to the thermal conductance. Figures 1 (a, b) and Figure 2 show the family of nine molecules investigated, which consist of dithiol-terminated oligo(phenylene-ethynylenes) (OPE) containing various central units. Molecule **2** differs from **3-9**, because the latter are conjugated, whereas the anthraquinone central unit of **2**, which contains pendant oxygens, is cross conjugated. Molecule **1**, which is the reduced form of **2** with hydrogens attached to the oxygens, is also fully conjugated

When a molecule is coupled to electrodes and the energy $E$ of an electron approaches the energy $\varepsilon_m$ of a molecular orbital, then the transmission coefficient $T(E)$ is approximated by the Breit-Wigner formula $T(E) = 4\Gamma_1\Gamma_2/[(E- \varepsilon_0)^2 +(\Gamma_1+\Gamma_2)^2]$, where $\Gamma_1$ and $\Gamma_2$ are the level broadenings due to contact with the left (1) and right (2) electrodes and $\varepsilon_0 = \varepsilon_m - \Sigma_1 - \Sigma_2$ is the resonance energy, obtained by shifting $\varepsilon_m$ by the real parts of the corresponding self-energies, $\Sigma_1$ and $\Sigma_2$. This expression is a consequence of constructive quantum interference and shows that $T(E)$ is a maximum when $E = \varepsilon_0$. In contrast, when a pendant group attached to a molecule creates a bound state of energy $\varepsilon$, which interacts with such a molecular orbital, then, as shown in ref [18,] the above expression is replaced by the modified formula

$$T(E) = 4\Gamma_1\Gamma_2/[(E- \varepsilon')^2 +(\Gamma_1+\Gamma_2)^2] \quad (1)$$

where $\varepsilon'= \varepsilon_0 + \alpha^2/(E- \varepsilon)$, where $\alpha$ is the coupling between the pendant group and the molecule. This expression describes the destructive interference which occurs when an electron resonates with a pendant bound state, because when $E= \varepsilon$, $\varepsilon'$ diverges and $T(E)$ vanishes. It is known[17-19,24,35] that pendant oxygens promote destructive quantum interference[47,48] due to the creation of a localized state in the central part of the molecule, whose effect is similar to that of a side group state[24,35], and create sharp anti-resonances near the Fermi energy $E_F$, leading to a large suppression in electrical conductance[17-23,47,48]. Consequently electrochemical switching[24] from **1** to **2** should produce a large decrease in the electrical conductance. It is therefore of interest to determine whether or not $S$ and $ZT$ can also be controlled by varying the oxidation state of **1** and **2**.

## 2. Theoretical Methods.

Transport properties were computed using the ab-initio code SMEAGOL[25,26] which employs the Hamiltonian provided by the density functional theory (DFT) code SIESTA,[27] in combination with the nonequilibrium Green`s function formalism. SIESTA uses norm-conserving pseudo-potentials and linear combinations of pseudo-atomic orbitals to construct the valence states. The calculations used a single zeta (SZ) basis set for the gold leads, which included the *s-* and *d-*orbitals in the valence to make sure the coupling to the sulphur atoms was correctly described. An energy cut-off of 200 Ry was chosen to define the real space grid, necessary to represent the density and the potential and to calculate the Hamiltonian and the overlap matrix elements. The local density approximation (LDA)[28] was employed to calculate exchange and correlation effects. The molecular coordinates were relaxed in the isolated molecule until the forces were smaller than 0.05 eV/Å. This force tolerance was sufficient to relax the structure of the molecules and achieve properly converged transport properties when the molecule was placed between the gold surfaces. Such relaxed coordinates were then included into the "extended molecule" without further relaxation (relaxation of the atomic coordinates inside the extended molecule gave essentially the same results)[31]. The "extended molecule" is represented by the dithiolated OPE compound bridged to five atomic gold layers on each side to account for specific properties of the contact region such as screening, adsorption geometry and molecular conformation. Each layer contained 18 atoms, i.e. 6 and 3 atoms along each lattice vector on the (111) surface. Only the gamma point was considered to perform the Brillouin zone integration along the perpendicular directions, which in case of conjugated molecules coupled via thiol groups to gold surfaces is enough to converge the results around the Fermi level[31,46]. One of the sulfurs was bound to a threefold hollow (H) site of the Au(111) surface (Au - S distance ~0.20 nm) and the other one to a gold adatom, which was contacted to the surface in a hollow site (Au - S distance ~0.24 nm) following the (111) direction of growth. In practice, the binding geometry can affect the electrical conductance, leading to high, medium and low conductance groups in break-junction experiments. This variation in the coupling to the electrodes affects the values of the parameters $\Gamma_1$ and $\Gamma_2$ of equation (1) and therefore affects the widths of Breit-Wigner resonances. On the other hand the parameter $\alpha$ of equation (1) is not affected by the coupling to the leads and therefore the width of the Fano resonance is relatively insensitive to the binding geometry.

Corrected positions of the Fermi energy relative to the HOMO to account for the DFT underestimation of the HOMO-LUMO gap were obtained by applying a scissor-type operator to the bare DFT results (SAiNT[29,30]) that moves the occupied and the unoccupied molecular levels downwards and upwards, respectively. This spectral adjustment was chosen to fit the position of such levels as obtained from a combination of experimental UPS, UV/VIS absorption and electrochemical data and has been verified through comparison with electrical measurements on molecules **1-9**[31,32]. The Hamiltonians derived in these arrangements are then used by SMEAGOL to calculate the transmission coefficients *T(E)*.

To calculate thermoelectric quantities[17] it is useful to introduce the non-normalised probability distribution *P(E)* defined by

$$P(E) = -T(E)\frac{\partial f(E)}{\partial E},$$

where *T(E)* is the transmission coefficient for electrons of energy *E* passing from one electrode to the other and *f(E)* is the Fermi-Dirac function, whose moments are denoted

$$L_i = \int dE\, P(E)(E - E_F)^i, \qquad (2)$$

where $E_F$ is the Fermi energy. The conductance, *G* is

$$G(T) = \frac{2e^2}{h} L_0, \qquad (3)$$

where *e* is the electronic charge, *h* is the Planck constant, and *T* is the temperature. The thermopower *S* is

$$S(T) = -\frac{1}{eT}\frac{L_1}{L_0}, \qquad (4)$$

the electronic contribution to the thermal conductance $\kappa_{el}$ is

$$\kappa_{el} = \frac{2}{h}\frac{1}{T}\left(L_2 - \frac{L_1^2}{L_0}\right), \qquad (5)$$

and the electronic contribution to the figure of merit $Z_{el}T$ is

$$Z_{el}T = \frac{1}{\frac{L_0 L_2}{L_1^2} - 1}. \qquad (6)$$

All our results were obtained using the above formulae. As an aside, for E close to $E_F$, if $T(E)$ varies only slowly with E on the scale of $k_B T$ then these expressions take the well-known forms:

$$G(T) \approx \left(\frac{2e^2}{h}\right) T(E_F), \qquad (7)$$

$$S(T) \approx -\alpha e T \left(\frac{d\, lnT(E)}{dE}\right)_{E=E_F}, \qquad (8)$$

$$\kappa \approx L_0 T G,$$

where $\alpha = (\frac{k_B}{e})^2 \pi^2/3$ is the Lorentz number. The latter expression demonstrates the "rule of thumb" that S is enhanced by increasing the slope of ln $T(E)$ near $E=E_F$.

To compute the phonon contribution $\kappa_p$ to the thermal conductance, we evaluated the expression

$$\kappa_p = \frac{1}{2\pi} \int_0^\infty d\omega\, \hbar\omega [df_B(\omega,T)/dT]\tau(\omega) \quad (9)$$

where $\tau(\omega)$ is the transmission coefficient for phonons of frequency $\omega$ and $f_B(\omega,T)$ is the Bose–Einstein distribution[33,34].

To compute $\tau(\omega)$ we placed molecules **1** and **2** between simple quasi one-dimensional gold leads shown in Figure 5and geometrically optimized these structures. The geometry optimization was carried out with the classical molecular dynamics code LAMMPS and for the force field parametrization we used the reaxFF force field. We applied open boundary conditions in all directions and for each atom, the tolerance of each force component during the optimization was 4.0e$^{-12}$eV/A. After geometry relaxation, we obtained the dynamical matrix of the system using finite differences to calculate the force derivatives with respect to x, y and z components of all atoms. More precisely, the elements of the dynamical matrix are calculated as $D_{I\alpha,J\beta} = -\frac{1}{\sqrt{m_I m_J}} \frac{F_{I\alpha}(r_J+he_\beta)-F_{I\alpha}(r_J-he_\beta)}{2h}$, where I and J are the atom indecies and $\alpha$ and $\beta$ are x,y or z. The $m_I$ and $m_J$ denote the mass of atoms I and J and $F_{I\alpha}(r_J \pm he_\beta)$ denotes the $\alpha$ component of the force acting on the atom I, when atom J is displaced along the $\beta$ direction by $\pm h$ from its equilibrium position ($r_J$), where h=0.01A. The forces acting on each atom were calculated again using LAMMPS with reaxFF force field. The transmission coefficient $\tau(\omega)$ is then computed by calculating the phonon Green's function connecting the left to the right gold leads, as discussed in ref[34].

The results are shown in Figure 6. The various "noisy" peaks are associated to the many molecular resonance frequencies. Higher than $\hbar\omega > 0.02eV$ there is no transport channel supported by the simple gold leads. The difference in the transmission coefficients shown in Figure 6 of the two structures shown in Figure 5 arise from the difference in their molecular structure and the slightly different binding geometries. However, as shown in Figure 7, despite these differences the thermal conductance of the two structures are of the same order of magnitude, reflecting the fact that the two molecules have similar rigidities.

## 3. Results

Figure 1 (a) shows a dithiol-terminated oligo(phenylene-ethyny-lene) (OPE) containing a hydroquinonecentral unit **1** connected via thiol anchor groups to gold electrodes, which mimic a STM tip and a surface. Figure 1 (c) (black curve) shows that the transmission coefficient $T(E)$ for electrons passing from one electrode to the other is rather featureless in the vicinity of the Fermi energy ($E - E_F = 0$). Electrochemically oxidising this molecule removes the hydrogens from the –OH groups to yield the cross-conjugated anthraquinone central unit **2** shown in figure 1 (b), which possess pendant oxygens. As expected, figure 1 (c) (green dashed curve) shows that pendant oxygens produce destructive interference in the vicinity of $E_F$, indicated by a sharp decrease in $T(E)$ near $E_F$. The destructive interference is due to the presence of a state localized in the central part of the molecule induced by the oxygens[24,35]. Such a state gives rise to an additional path through which electrons can pass. This path and additional paths associated with other states close in energy give rise to interference effects that suppress the conductance near the Fermi level. Since the electrical conductance $G$ is given by $G = G_0 T(E_F)$, where $G_0$ is the quantum of conductance, one expects that the conductance of **2** to be significantly suppressed. This prediction has been confirmed experimentally only very recently[31,32,35,36]. The passivation of the oxygens removes two electrons from the system and moves the localised state away from the Fermi level, cancelling the interference effect near $E_F$ and thus increasing the conductance.

For **1** and **2**, figure 3 shows the variation of $G$, $S$, $\kappa_{el}$ and $Z_{el}T$ with temperature and demonstrates that at room temperature, the redox switching from **1** to **2** produces a three-orders-of-magnitude suppression of $G$ and $\kappa_{el}$ and a large enhancement of $S$ and $Z_{el}T$. This enhancement is ultimately associated with the appearance of the anti-resonance near $E_F$,[17,20,49,50,51] because, as noted in equation (8), at low temperatures $S$ is proportional to the slope of the logarithm of $T(E)$, evaluated at $E = E_F$. The Seebeck coefficient can therefore be dramatically enhanced by using antiresonances (or transmission nodes) that increase the logarithmic derivative. The sharper the antiresonance and the closer it is to the Fermi level, the larger $S$. Such sharpness could in principle be implemented without sstrongly suppressing the conductance by placing the position of the node as close as possible to a transmission resonance, as shown in figure 3 of reference [18].

To demonstrate that the large room-temperature value of $S = 80\ \mu V/K$ obtained for **2** is unprecedented and much higher than those obtained for other OPE-based molecular wires, figure 4 (a) shows plots of $S$ for the remaining seven members of this family of molecular wires, shown in figure 2. At 300K, the thermopowers of all other molecules are approximately a factor of two or less than that of **2**. Literature values for S obtained experimentally are also significantly lower than that of **2**. For example, recently-measured values of S at room temperature include 8.7, 12.9 and 14.2 $\mu V/K$ for 1,4-benzenedithiol (BDT), 4,4'-dibenzenedithiol, and 4,4''-tribenzenedithiol in contact with gold respectively[12], -1.3 to 8.3 $\mu V/K$ for the benzene-based series of benzene-dithiol (BDT), 2,5-dimethyl-1,4-benzenedithiol (BDT2Me), 2,3,5,6-tetrachloro-1,4-benzenedithiol (BDT4Cl), 2,3,5,6-tetraflouro-1,4-benzenedithiol (BDT4F) and BDCN[11,13], 7.7 to 15.9 $\mu V/K$ for the series BDT, DBDT, TBDT and DMTBDT[16], -12.3 to 13.0 $\mu V/K$ for a series of amine-Au and pyridine-Au linked molecules[37] and -8.9 to -33.1 $\mu V/K$ for fullerene-based single-molecule junctions[15,38,39] . The thermopower of molecule **2** exceeds all of these values. It also exceeds theoretical predictions of $S$, provided corrections to DFT are implemented to yield agreement with experiment[40,41].

In practice the task of measuring the thermal conductance $\kappa$ is non-trivial and experimental values of $ZT$ are usually not reported in the literature. Figure 4 (b) demonstrates that the electronic figure of merit $Z_{el}T$ of **2** is also much higher than that of molecules **1** and **3-9** and therefore, if the phonon contribution $\kappa_p$ to the thermal conductance can be reduced to a value below that of $\kappa_{el}$, this would form the basis of an attractive thermoelectric converter. As stated above, the basis is a sharp antiresonance that is close enough to the Fermi level, so that the logarithmic derivative is greatly enhanced.

To estimate the phonon contribution, we have calculated the phonon transmission coefficients $\tau(\omega)$ for the structures shown in figure 5. These are shown in figure 6 and the resulting temperature dependences for $\kappa_p$ obtained by evaluating equation 9 are shown in figure 7.

This illustrates that for molecules **1** and **2** attached to gold electrodes, the ratio $\kappa_p/\kappa_{el} \gg 1$ and therefore $ZT \ll Z_{el}T$. To remedy this suppression, strategies for decreasing the role of parasitic phonons should be implemented. Possible approaches include utilising electrodes made from materials with lower thermal conductance than gold, nanostructuring the electrodes to introduce phonon band gaps, or utilising disordered electrodes which maximise phonon scattering or even introduce phonon localisation[42,43,44].

## 4. Summary

In summary, we have demonstrated that redox control of cross-conjugated molecules[45] is not only a method for tuning the electrical conductance, but also is a viable strategy for enhancing the thermoelectric performance of molecular junctions. Thermopower, unlike electrical conductance, is an intrinsic quantity, which does not scale with the surface area of a monolayer of molecules or with the contact configuration[29]. Therefore the above results for single molecules point the way towards engineering the thermoelectric properties of thin films. The idea of utilising destructive quantum interference to yield enhanced thermoelectric properties was discussed in [17] and observation that pendant oxygens lead to anti-resonances close to the Fermi energy was highlighted in [18,19]. By employing a corrected DFT Hamiltonian, which has been benchmarked against experiments in Ref. [31], the above calculations demonstrate that these two effects combine in molecules **1** and **2** to yield enhanced electrochemically-switchable values of $S$ and $Z_{el}T$, which are competitive with the best available organic materials. Such values of $S$ and $Z_{el}T$ are highly unusual, as demonstrated by comparison with the family of related molecules **3-8** and comparison with recent experiments. For the future, it will be of interest to examine the thermoelectric properties of related families of molecules with pendant oxygens, such as the fluorene-based example of Ref. [19].

**Figures**

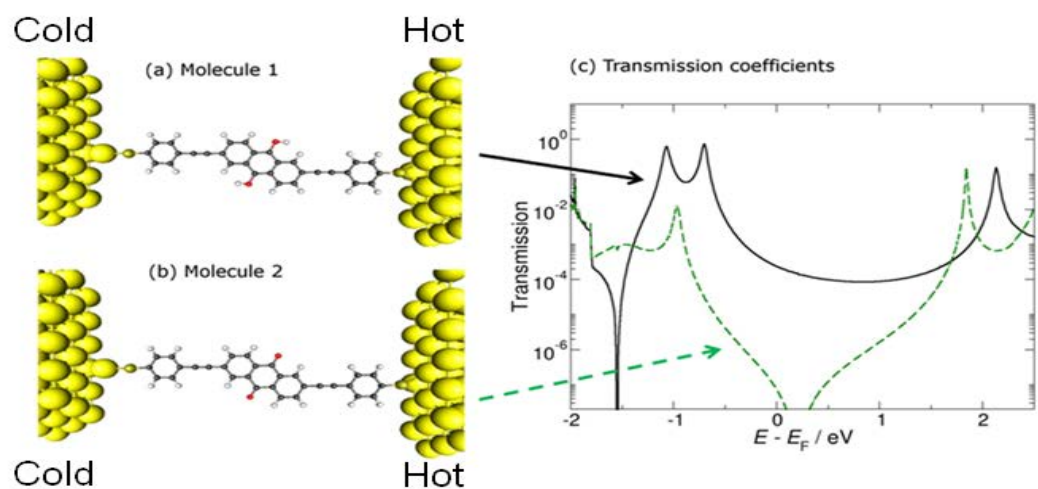

**Figure 1.** (1a) A dithiol-terminated oligo(phenylene-ethyny-lene) (OPE) containing a hydroquinonecentral unit **1** connected via thiol anchor groups to gold electrodes. (1b) The cross-conjugated anthraquinone central unit **2**. (1c) Transmission coefficients $T(E)$ for **1** (continuous curve) and **2** (dashed curve).

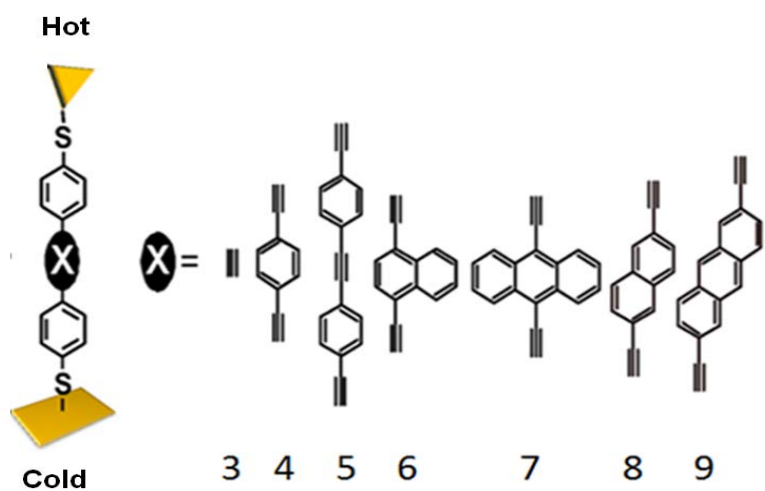

**Figure 2.** Seven OPE-based wires with different lengths and central units, labeled **3**-**9**.

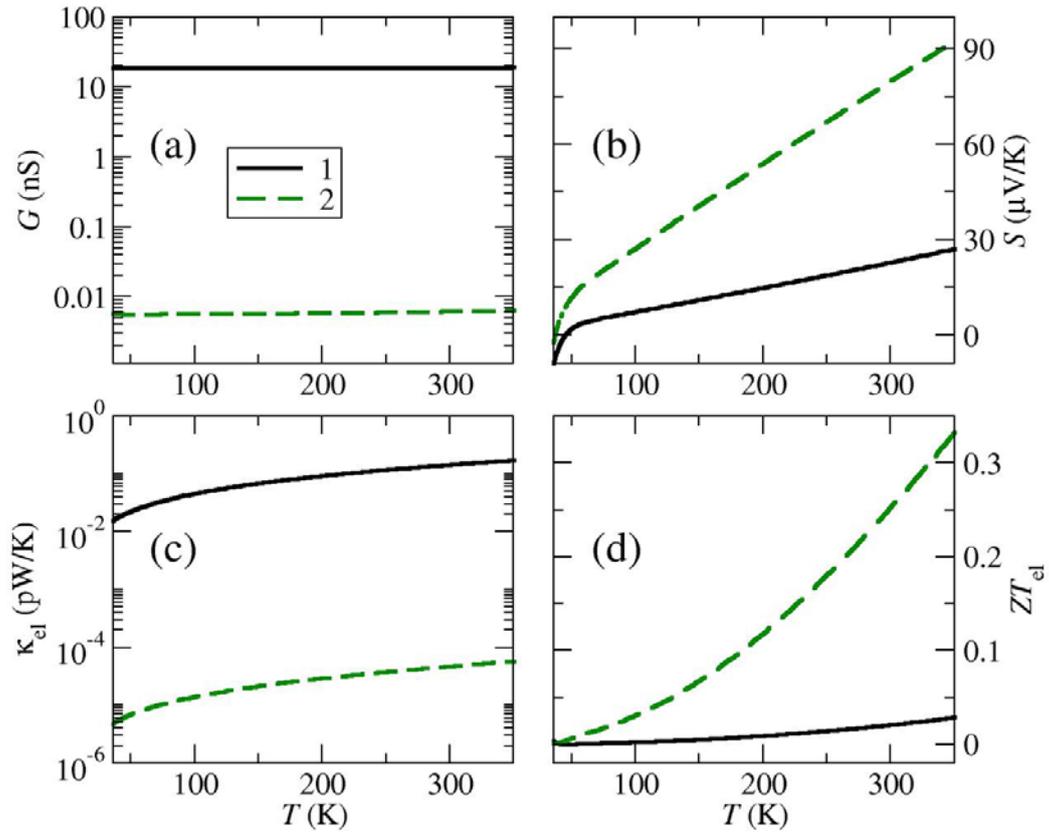

**Figure 3.** (2a), (2b), (2c) and (2d) show, respectively, the variation of the electrical conductance ($G$), the Seebeck coefficient ($S$), the electronic thermal conductance ($\kappa_{el}$) and the electronic figure of merit ($ZT_{el}$) with temperature for molecules **1** and **2**.

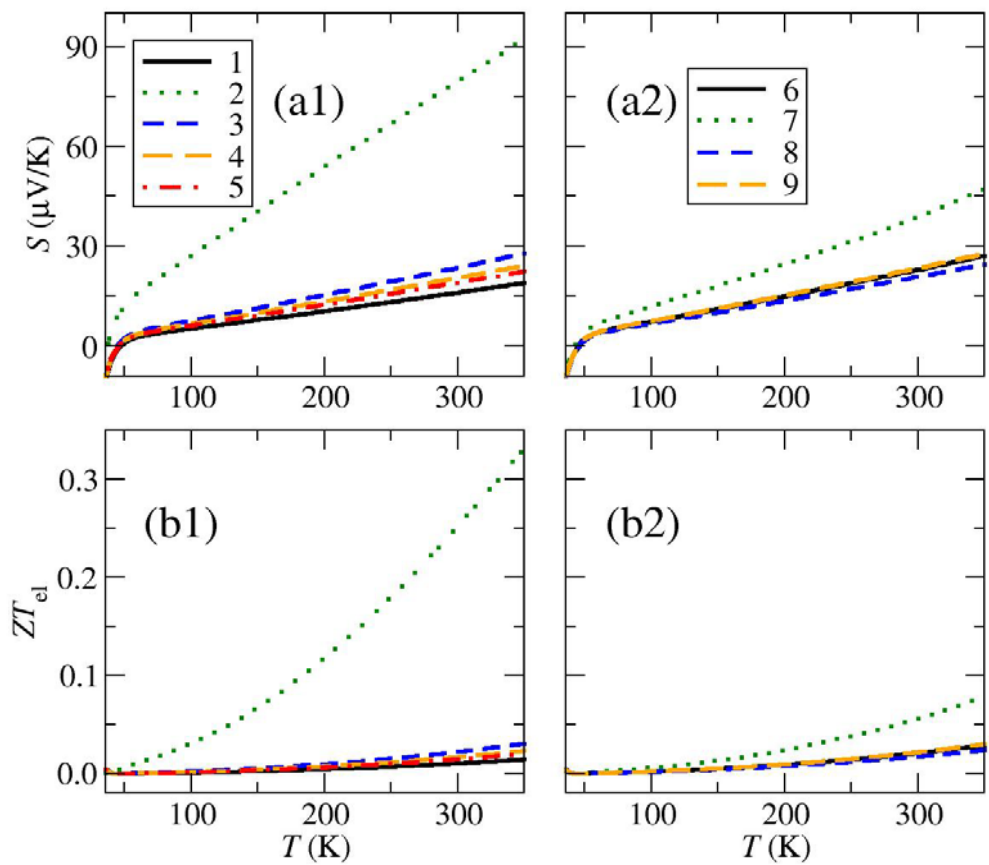

**Figure 4**

**Figure 4.** (a) Variation of $S$ with temperature for the nine OPE-based wires of Figs. 1 and 3. (b) Variation of $ZT_{el}$ with temperature for the same nine wires.

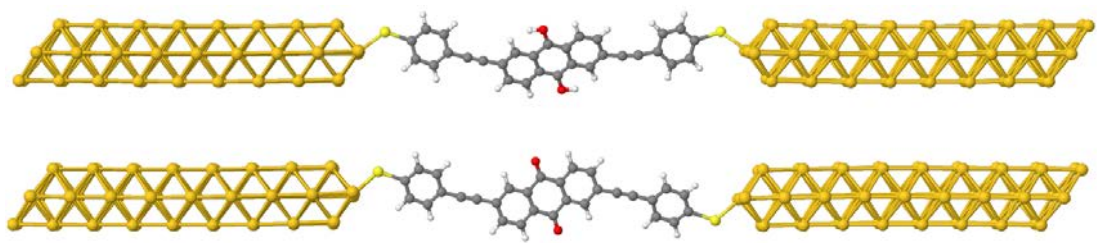

**Figure 5.** Optimized geometries of molecules **1** and **2** connected to simple gold leads. The lead has a cross section of 7 atoms. The geometry optimization is performed with reaxFF forcefield by using LAMMPS.

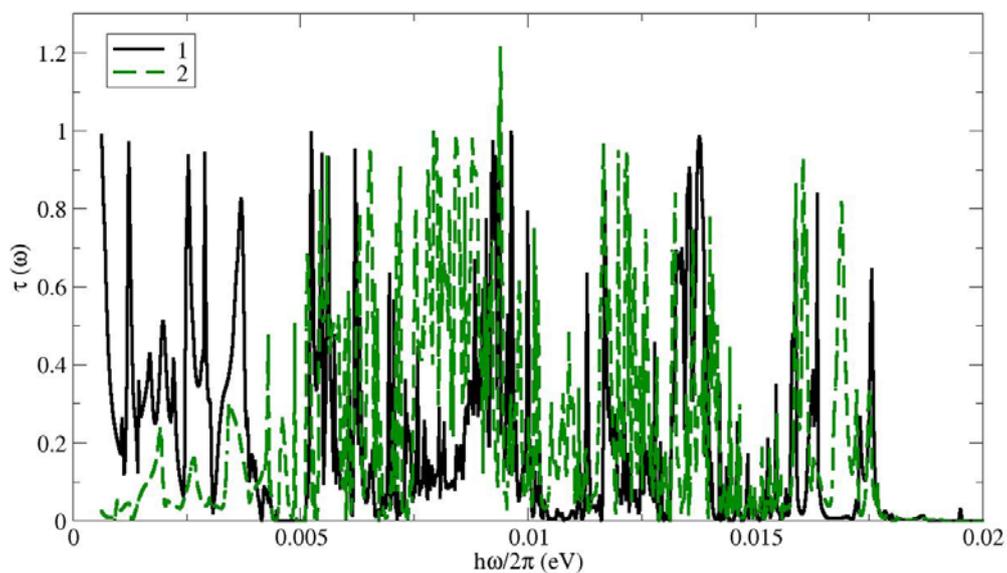

**Figure 6.** The phonon transmission coefficients for the molecular systems **1** (black) and **2** (green) shown in figure 5.

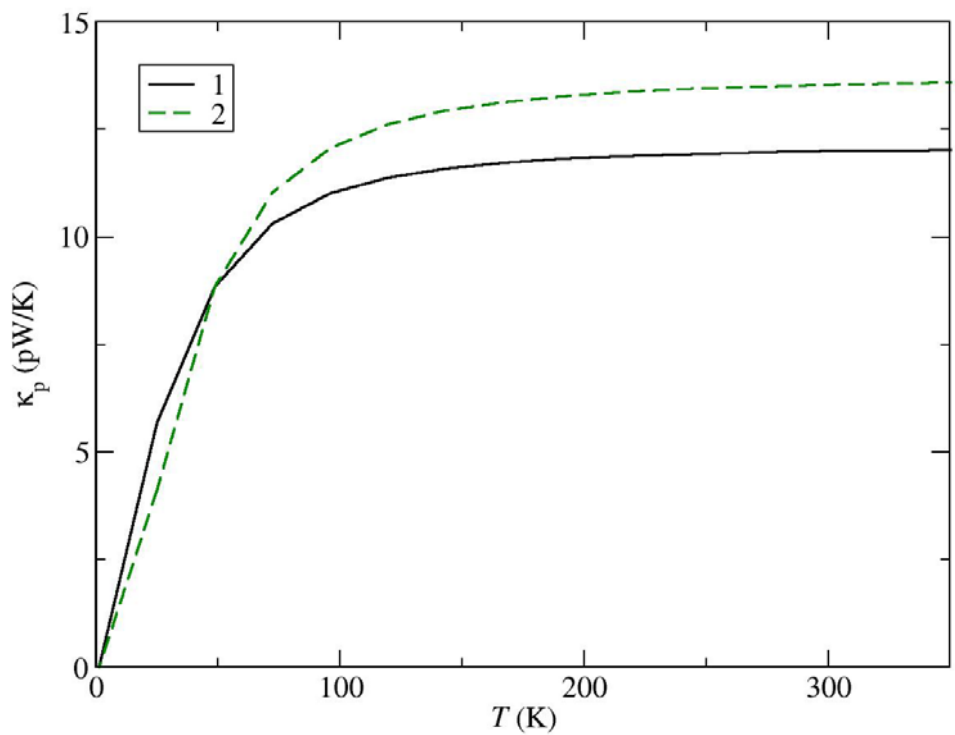

**Figure 7.** Phonon contribution of the thermal conductance for the molecular systems shown in figure 5.

AUTHOR INFORMATION

**Corresponding Author**

*E-mail c.lambert@lancaster.ac.uk.

**Author Contributions**

The authors contributed equally to this work.

NOTES

The authors declare no competing financial interest.

ACKNOWLEDGEMENTS

This work was supported by the European Union (FP7) ITN NANOCTM and by the EPSRC. VMGS thanks the Spanish Ministerio de Economía y Competitividad for a Ramón y Cajal fellowship (RYC-2010-06053). TW acknowledges support from SNF through grant 200020_144471/1.